%% file: main.tex
\documentclass[a4paper,11pt]{article}
\usepackage{aaskaiid}
\usepackage{wrapfig}
\usepackage{tcolorbox}
\usepackage{orcidlink}

\title{Galactic Hubs: New Insights and SKA View of AGN Feedback in Galaxy Groups}
\ShortTitle{SKA view of AGN feedback in galaxy groups}

\author[1]{Thomas Pasini\orcidlink{0000-0002-9711-5554}}
\ShortName{T. Pasini et al.}
\author[1]{Marisa Brienza\orcidlink{0000-0003-4120-9970}}
\author[2,3]{Christopher J. Riseley\orcidlink{0000-0002-3369-1085}}

\affiliation[1]{Istituto di Radioastronomia IRA-INAF, Via Gobetti 101, 40129 Bologna, Italy}
\emailAdd{thomas.pasini@inaf.it}
\affiliation[2]{Astronomisches Institut der Ruhr-Universit\"{a}t Bochum, Universit\"{a}tsstra{\ss}e 150, 44801 Bochum, Germany}
\affiliation[3]{Ruhr Astroparticle and Plasma Physics Center (RAPP Center), 44780 Bochum, Germany}

\abstract{In the last two decades, significant effort has been devoted to understanding the physical processes involved in galaxy evolution. In this context, AGN feedback has emerged as the leading explanation for several longstanding astrophysical questions, including the origin of scaling relations between supermassive black holes and their host galaxies, as well as the low and strongly mass-dependent efficiency with which baryons are converted into stars compared to the expectations of simple cooling models.
Galaxy clusters have been particularly put under the lens, thanks to the combination of high-resolution X-ray and radio observations which allowed to investigate the link between thermal and non-thermal emission, their connection to feedback and its effects on BCGs and member galaxies. On the other hand, research in the lower-mass regime of galaxy groups remains limited. This is mostly due to observational limitations in the high-energy regime, which prevents the study of the non-thermal emission in the context of its surrounding environment. Nevertheless, this is a key topic for galaxy evolution, as the vast majority of radio-loud AGN in the nearby Universe are hosted in galaxy groups. 

Recently, modern radio facilities such as LOFAR, uGMRT and MeerKAT have obtained interesting results in this regime. Examples include signatures of overheating caused by AGN, groups with largely evacuated environments due to powerful and prolonged feedback activity, and remnant plasma displaced by large-scale sloshing motions, which all demonstrate how the evolution of groups and AGN can influence each other.
This chapter reviews these advances while identifying gaps in our understanding. We will explore how SKA promises to close these gaps and revolutionise the field. With its unparalleled combination of high resolution and sensitivity across a wide frequency range, SKA will enable detailed spectral analysis on various physical scales, offering new insights into the complex interactions that shape galaxy group evolution. Moreover, its unique capabilities, in synergy with state-of-the-art and upcoming surveys carried out by new instruments (e.g. eROSITA, 4MOST, Euclid) will facilitate the construction of unprecedentedly large samples of groups, dramatically enhancing the statistical power of future studies.}

\begin{document}
\include{journal-names}
\maketitle
\thispagestyle{plain}     
                            

\clearpage                   
\pagenumbering{arabic}      
\setcounter{page}{1}   

\section{Introduction}

Over the last couple of decades, a growing number of studies have demonstrated that the mass function in the current Universe is dominated by halos with masses in the range $10^{13} < M_\odot < 10^{14}$ \citep[see e.g.,][]{Tinker_2008}, which corresponds to the typical regime of galaxy groups. The existence of galaxy groups and of their more massive counterparts, galaxy clusters ($M > 10^{14} M_\odot$), reveals that the Universe is not perfectly homogeneous. Instead, it emerged from the Big Bang with primordial overdensities that, under the influence of gravity, gradually evolved into the large-scale structures we observe today \citep[e.g.,][]{White_1978}.

This has important implications for cosmology, as the number, distribution, and growth of galaxy groups and clusters are tightly linked to the underlying cosmological parameters and to the physics governing the evolution of the Universe \citep{Rosati_2002}. Consequently, cosmological models must be able to accurately predict and reproduce the observed properties and spatial distribution of these systems.

If gravitational processes were the only drivers of their formation and evolution, the global properties of the Intra-Cluster Medium/Intra-Group Medium (ICM/IGrM) of galaxy clusters and groups, such as mass, temperature and X-ray luminosity, should follow tight scaling relations. However, observational studies have revealed significant deviations from these self-similar predictions \citep[e.g.,][]{Markevitch_1998}, suggesting that non-gravitational processes, and particularly feedback from active galactic nuclei (AGN), play a crucial role in shaping the thermal and dynamical state of the ICM and IGrM \citep{skagitti}. Understanding the impact of such processes is essential for interpreting the discrepancies between theoretical predictions and observations, and ultimately for improving constraints on cosmological models.

In this context, galaxy groups are particularly important as they host more than half of all galaxies in the Universe and contain a large fraction of its baryonic content \citep{Eke_2006}. This makes them ideal laboratories for studying galaxy evolution and the interplay between galaxies and their surrounding environment. Radio observations are particularly important for this, as the synchrotron non-thermal emission often detected in these environments usually interacts with the IGrM, and heavily affects the evolution of the host system.

In this chapter, we review recent advances in the study of galaxy groups made through radio observations, which were possible thanks to the advent of SKA pathfinder and precursor facilities. We discuss how these observations have begun to reveal the presence and impact of non-thermal components—such as relativistic particles and magnetic fields—within the group environment. Finally, we explore how the upcoming capabilities of SKA will revolutionize our understanding of galaxy groups by enabling high-sensitivity, high-resolution studies of their radio emission. This will provide crucial insights into the physical processes driving AGN feedback, the origin and evolution of cosmic magnetism, and the role of non-thermal phenomena in structure formation and evolution.

\section{Cosmological context}

The importance of galaxy groups for cosmology has been well known for more than 50 years. \citet{Press_1974} first developed an analytical framework to predict the number density of gravitationally bound halos as a function of mass and cosmic time. In their models, the high-mass ($>10^{14} M_\odot$) end is dominated by an exponential term which results in a steep decline in terms of number of massive halos. On the other hand, the lower-mass range roughly follows a power-law $dn / dM \propto M^{-2}$. Galaxy groups start to form around $z \sim 2$, and become increasingly common at lower redshifts.

\begin{figure}[h!]
    \centering
	\includegraphics[height=10cm, width=11cm]{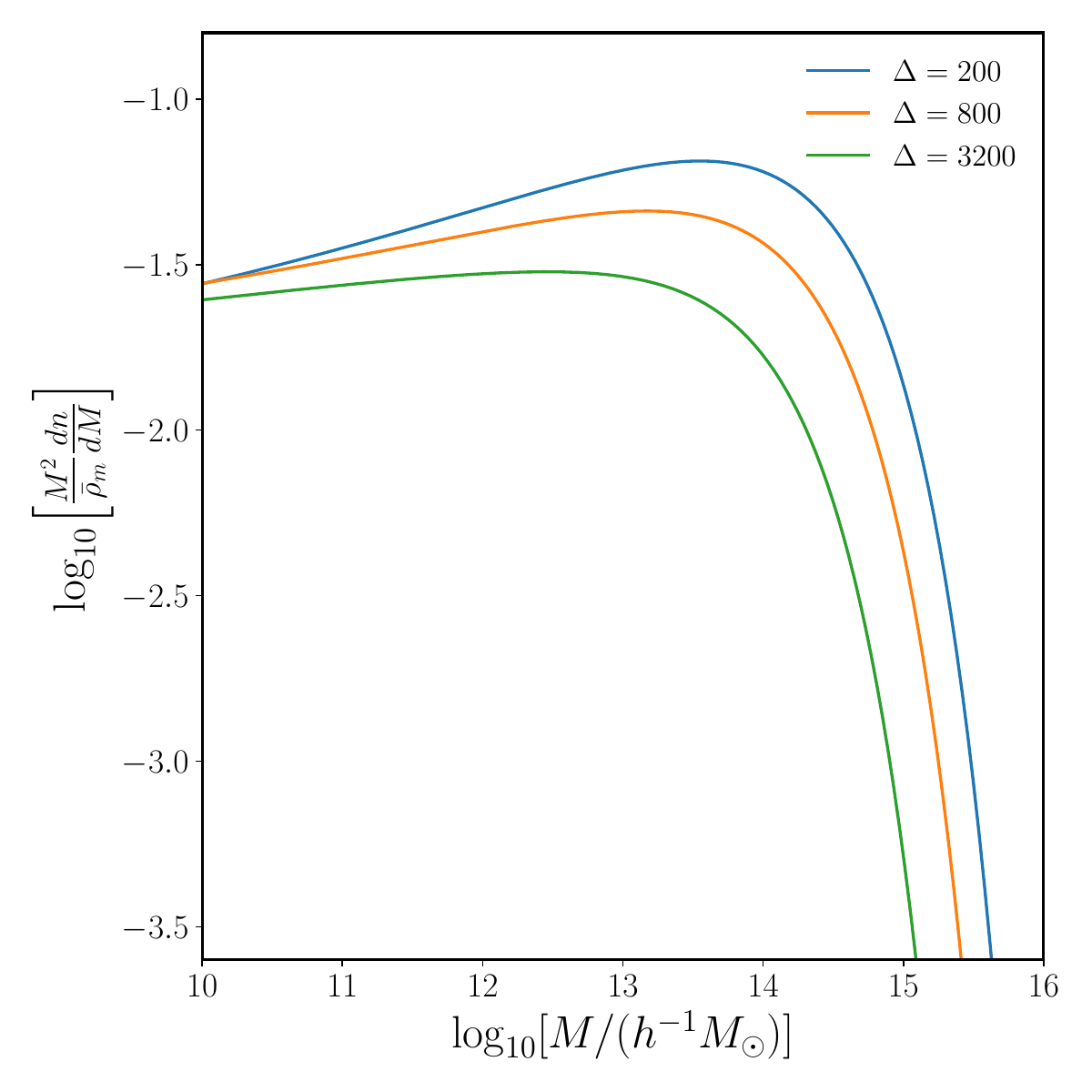}
    \caption{Measured mass function for all simulations included in \citet{Tinker_2008} which adopt a WMAP1 cosmology. The three curves show the mass function for different values of $\Delta$, which is the highest overdensity for which the function can be directly measured in these simulations.}
    \label{fig:massfunction}
\end{figure}

In Fig. \ref{fig:massfunction} we show the mass function at $z = 0$ for halos with masses between $10^{10}$ and $10^{16}$ M$_\odot$, adapted from \citet{Tinker_2008}. The three curves represent different values of $\Delta$, which parametrizes the highest overdensity for which the mass function can be measured directly in simulations adopting a WMAP1 cosmology \citep{Spergel_2003}. While we refer to \citet{Tinker_2008} for additional details, it is clear that the curves show a peak around $M \sim 10^{14}$ M$_\odot$. As a result, at $z=0$ galaxy groups are the dominant collapsed structures in the Universe.
 
Following hierarchical structure formation, galaxy groups constitute the building blocks of galaxy clusters. In particular, according to \citet{White_1978}, groups act as intermediate nodes through which galaxies assemble before becoming part of more massive clusters. Approximately half of all galaxies today have evolved within group environments for at least part of their history \citep{Eke_2006}. \citet{White_1978} first presented a cosmological formation model where galaxies are formed through radiative cooling of baryons in halos that were previously generated due to dark matter (DM) gravitational collapse. In this self-similar scenario where gravitational dynamics is the only factor, precise scaling relations are expected to exist between group/cluster properties such as mass, temperature, entropy and luminosity. However, observations clearly show that these properties significantly deviate from these simple relations, suggesting that additional mechanisms are also at play \citep[e.g.,][]{Markevitch_1998, Arnaud_2005, Ettori_2015, Lovisari_2020}. Furthermore, without introducing radiative processes, simulations are not able to reproduce the observed thermodynamic properties of groups (see e.g. \citealt{Ascasibar_2006, Springel_2006}, or the comprehensive review of \citealt{Eckert_2021}). We will return to this in the following sections.

The vast majority of the baryonic budget of the Universe is known to reside in the atmospheres of groups and clusters \citep[e.g.,][]{Fukugita_1998}, with only a small fraction ($<10 \%$) that has condensed into stars. In fact, observational evidence over the past two decades has revealed a significant mismatch between the baryon content measured in galaxy groups and the universal baryon fraction inferred from cosmological observations, such as those provided by \textit{Planck}. This discrepancy becomes more pronounced at lower masses, suggesting that the processes governing baryon retention and distribution are strongly mass-dependent. 

\citet{Giodini_2009} combined X-ray data from XMM-Newton with weak lensing mass estimates to derive the baryonic mass fraction (including both hot gas and stars) for a large sample of galaxy groups and clusters. Their results confirmed that, while massive clusters tend to approach the cosmic baryon fraction as predicted by $\Lambda$CDM cosmology and measured by \textit{Planck}, galaxy groups consistently fall short of this value. The deficit increases with decreasing halo mass. Furthermore, recent Sunyaev–Zel'dovich (SZ) observations from  \citet{Planck_2014} show that systems with masses $M_{500} \lesssim 10^{14} M_\odot$ usually retain only 50–70\% of the baryons expected based on the cosmic mean. The origin of these missing baryons remains an open question, but evidence suggests they may be located in a warm–hot intergalactic medium (WHIM) diffuse gas at temperatures of $10^5$–$10^7$ K, or have somehow been expelled from the group's potential well.

\begin{figure}[t!]
    \centering
	\includegraphics[scale=0.35]{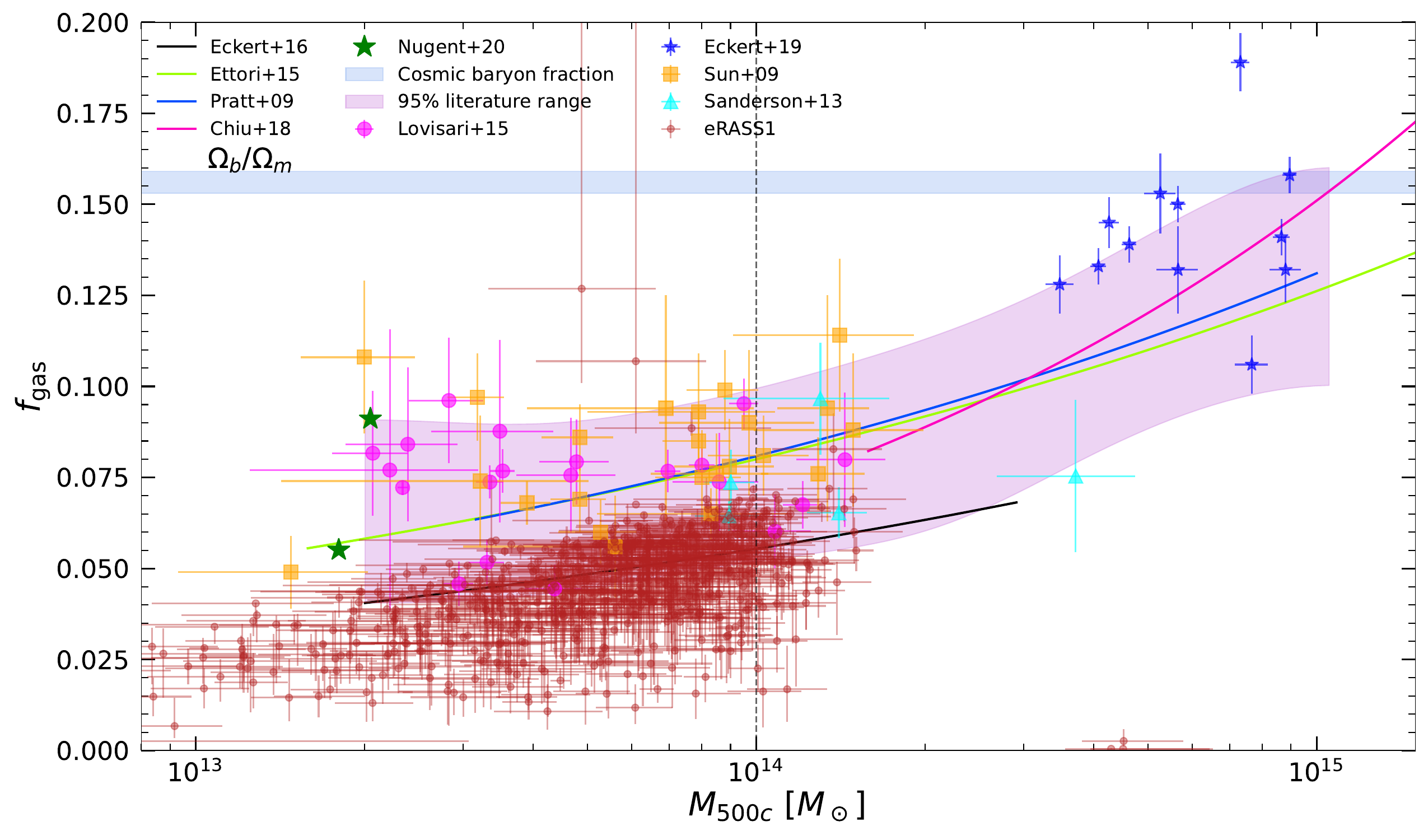}
    \caption{The figure, adapted from \citet{Eckert_2021}, shows a compilation of literature measurements of the hot gas fraction at $R_{500}$ in galaxy groups and clusters as a function of $M_{500}$. We additionally include recent eRASS galaxy group measurements from \citet{Bulbul_2024}, shown as red circles. The vertical dashed line indicates the conventional mass threshold separating the group and cluster regimes. The light blue shaded band marks the cosmic baryon fraction, while the purple shaded region represents the 95\% range of published measurements.}
    \label{fig:baryon_fraction}
\end{figure}

The observed differences with respect to self-similar predictions are highlighted in Fig. \ref{fig:baryon_fraction}, which shows the hot gas fraction measured at $R_{500}$ as a function of $M_{500}$ for samples of galaxy clusters and groups. The bulk of the plot is from \citet{Eckert_2021}, and we show in addition the eROSITA measurements\footnote{Only galaxy groups are shown.} from \citet{Bulbul_2024}. In massive clusters, the measured gas fraction is consistent (or slightly lower) with the cosmic baryon fraction predicted assuming self-similar structure formation. Moving to lower halo masses, the difference becomes significantly more pronounced and points to a strong impact of non-gravitational processes in the evolution of galaxy groups. In fact, new measurements from eROSITA point to an even more significant gas evacuation in this low-mass regime, with respect to what previously found with other instruments \citep{Bahar_2024, Grandis_2024, Siegel_2025}.  

These discrepancies in terms of scaling relations and baryon content highlight the need for additional non‐gravitational processes, which must be able to regulate the thermodynamic state and baryonic assembly of group‐scale halos. In the following section, we focus on how active galactic nucleus (AGN) feedback provides a compelling solution to these challenges.

\section{AGN feedback mechanisms in galaxy groups}

The term AGN feedback encompasses the processes by which a central (galactic) supermassive black hole injects energy and momentum into the gas of the host halo, thereby altering its thermodynamic state and evolution. In massive galaxy clusters, this mechanism has proved to be able to re-heat the ICM, preventing catastrophic cooling and powerful gas inflows towards the cluster centre as originally predicted by pure cooling flow models \citep[e.g.,][]{Fabian_1994}. Furthermore, feedback can take either a \textit{negative} or a \textit{positive} form \citep{Zinn_2013}, depending on the impact that the AGN energy release has on the surrounding gas. In the former, the injection of energy and momentum through radiation, winds, or relativistic jets acts to heat, ionize, or expel the cold gas reservoir, thus preventing further cooling and effectively suppressing star formation within the host galaxy \citep[e.g.,][]{Morganti_2017}. In contrast, the \textit{positive} form occurs when the AGN-driven outflows or jets interact with the interstellar or intragroup medium in such a way as to compress the gas, increasing its density and potentially triggering localised episodes of star formation \citep[e.g.,][]{Maiolino_2017}.

\begin{figure}[t!]
    \centering
	\includegraphics[scale=0.265]{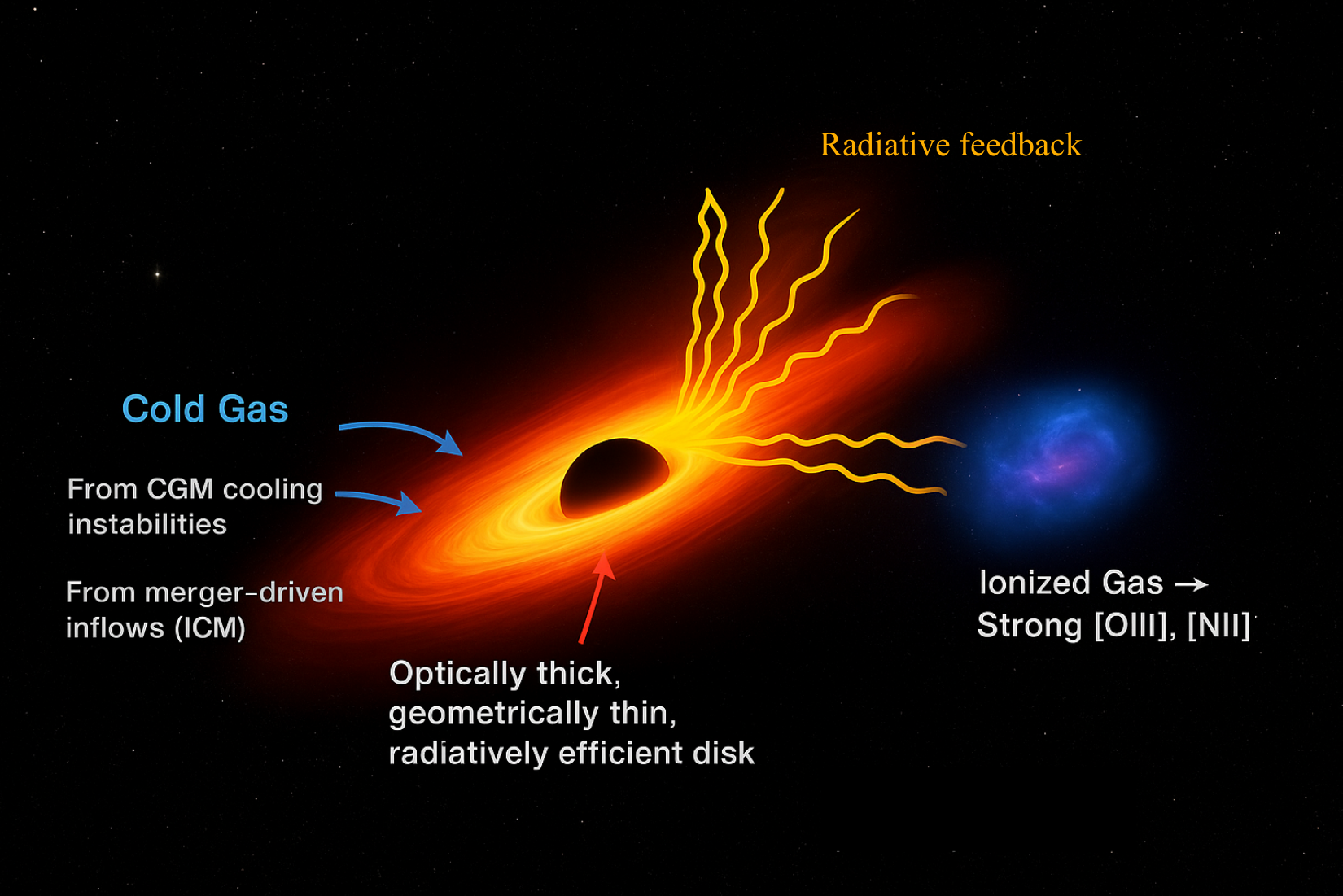}
        \includegraphics[height=4.95cm, width=7.5cm]{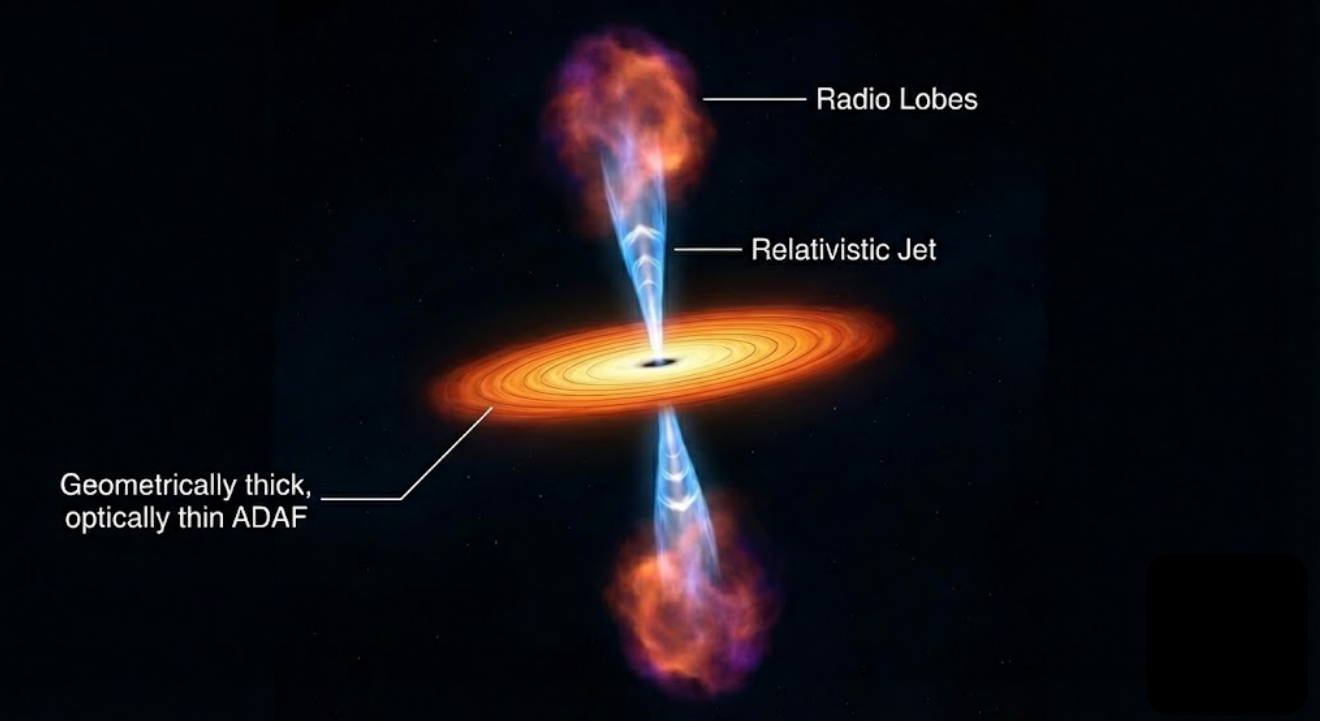}
    \caption{Artistic impression of the two main AGN modes. \textit{Left}: radiative mode, dominated by accretion of cold gas and mediated by an optically thick, geometrically thin disk. \textit{Right}: mechanical mode, mediated by ADAF onto an optically thin, geometrically thick disk.}
    \label{fig:feedback}
\end{figure}

AGN activity is typically divided into two main modes, which we show for clarity in Fig. \ref{fig:feedback} through an artistic impression. In the “radiative” or “quasar” mode, accretion onto the SMBH proceeds through an optically thick, geometrically thin, and radiatively efficient disk \citep{Shakura_1973}, and feedback is dominated by a strong radiation field, although some quasar-mode AGN are also known to host jets. In this regime the SMBH, which usually shown high accretion rate, is often fueled by cold gas, either from cooling instabilities in the circum-galactic medium (CGM) or by merger-driven inflows within the ICM \citep[e.g.,][]{Gaspari_2013, Gaspari_2017, Voit_2018, Hardcastle_2018}. Observationally, quasar-mode AGN are characterized by optical spectra showing strongly ionized gas and prominent high-excitation emission lines (e.g. [OIII], [NII]), and are commonly classified as High-Excitation Radio Galaxies (HERGs).

The second mode, often labeled “radio” or “jet” mode, operates when the SMBH slowly accretes via a radiatively inefficient, optically thin, and geometrically thick disk through advection‐dominated accretion flow (ADAF; \citealt{Narayan_1995}). In this configuration, the AGN lacks the strong high‐excitation optical lines characteristic of quasar‐mode systems and instead converts the bulk of its accretion power into kinetic energy \citep{Merloni_2007}. This kinetic output is funneled into kpc-to-Mpc relativistic jets and lobes, which can drive shocks into the ICM and inflate radio bubbles which, in turn, excavate cavities in the surrounding medium \citep[e.g.,][]{Oei_2024}. This mode is usually associated with Low‐Excitation Radio Galaxies (LERGs). Nevertheless, it is worth noting that the distinction between radiative/quasar and radio/jet modes should be considered as a phenomenological and simplified scheme: in reality, there is substantial overlap, and individual AGN can display both strong radiative output and powerful jets, or transition between the two regimes over time. Nevertheless, through these processes the jets can heat, displace, and mix the ICM, thereby preventing runaway cooling and regulating the long‐term thermal balance of cluster‐scale halos \citep{Best_2012}.

Modern simulations that cover galaxy evolution have started to widely implement prescriptions for AGN feedback. This is usually done by defining the feedback power as $\dot{E}_{AGN} = \varepsilon \dot{M}_{BH}c^2$, where $\varepsilon = \varepsilon_r \dot \varepsilon_f$ is an efficiency term which is the product of radiative efficiency ($\varepsilon_r$) and feedback efficiency ($\varepsilon_f$). The former represents the energy radiated away from accretion onto the SMBH, and is usually assumed to be $\varepsilon_r \sim 10\%$ \citep{Shakura_1973, Oppenheimer_2021}. The latter is the fraction of this energy that gets conveyed to the gas, either via thermal conduction and/or kinetically, and is often calibrated using the observed correlation between SMBH mass and halo mass, resulting in $\varepsilon_f \sim 0.15\%$. A number of simulations (e.g. Illustris, EAGLE, see \citealt{Vogelsberger_2014, Schaye_2015}) have also implemented dual AGN modes \citep[e.g.,][]{Sijacki_2007}, with a quasar mode and a radio mode that are triggered based on the efficiency of the accretion onto the SMBH. It is worth noting that jets, which are extremely complicated to model and are therefore hardly included in most of these simulations, have only been recently implemented in SIMBA \citep{Thomas_2021} and COLIBRE \citep{Husko_2025}. These prescriptions have allowed simulations to (more) accurately reproduce the observed properties of galaxies and of larger-scale environments such as groups and clusters, which would have not been possible without the inclusion of feedback.

Indeed, to date AGN feedback is the most successful mechanism for resolving several long‐standing problems in galaxy formation. It naturally establishes the observed correlations between supermassive black hole mass and host galaxy properties \citep[e.g.,][]{Magorrian_1998, Haring_2004} by linking black hole growth to the depth of the surrounding potential well. Without feedback, numerical models overproduce stars, converting more than 20\% of baryons into stars, whereas observations indicate stellar mass fractions below 10\% \citep[e.g.,][]{Bower_2006,Croton_2006}. Indeed, quasar-mode AGN at high redshifts are known to play a key role in suppressing star formation through powerful radiatively-driven winds \citep[e.g.,][]{DiMatteo_2005, Harrison_2017}. On the other hand, the radio mode is considered essential to prevent galaxies, and especially Brightest Cluster Galaxies (BCGs) and Brightest Group Galaxies (BGGs), from reigniting star formation. The mechanical AGN output is in fact able to compensate for the radiative losses of the ICM/IGrM, and in this way maintain the gas in a quasi-stable thermal equilibrium \citep[e.g.,][]{McNamara_2007, Fabian_2012, Heckman_2014}.
Therefore, by coupling mechanical and radiative energy to their surroundings, AGN are able to shape black hole growth, the star‐forming properties of their host galaxies, and the larger-scale environments where the galaxy is located.

However, while AGN feedback has been extensively studied in massive clusters \citep[e.g.,][]{McNamara_2007, McNamara_2012, Gitti_2012}, investigations in the galaxy group mass regime are still lacking. This is mainly because, in the X-ray band, signatures of AGN feedback such as cavities, shocks and cold fronts are more easily observable in bright, massive systems, rather than in smaller groups. Indeed, the combination of multi-wavelength data, especially from the X-ray and radio bands, has been essential to study how feedback operates in clusters. However, it has become clear that galaxy groups cannot be left out of the picture, or simply be considered as smaller versions of clusters. In groups, the shallower gravitational potential amplifies the impact of even moderate AGN outbursts. By periodically depositing mechanical and thermal energy in the group environment, AGN feedback can raise central gas entropy, flatten self-similar scaling relations, expel or redistribute baryons, and arrest catastrophic cooling over duty cycles of 10$^7$–10$^8$ yr \citep[see comprehensive review of][]{Eckert_2021}, thus resolving the deviations in both scaling laws and baryon content described in the previous section.

\section{Statistical studies of AGN feedback in galaxy groups}
\label{sec:statistics}

Over the last decade, radio astronomical observations have been revolutionised by the advent of the SKA pathfinders and precursors, such as the upgraded Jansky Very Large Array (JVLA), the upgraded Giant Metrewave Radio Telescope (uGMRT), the Low-Frequency Array (LOFAR), the Murchison Widefield Array (MWA), the Australian Square Kilometre Array Pathfinder (ASKAP), and MeerKAT. These instruments have delivered key results that have significantly advanced our understanding of radio galaxies and their environments, including galaxy groups.
In particular, the high sensitivity and resolution of most of these facilities have enabled survey observations of wide sky areas. These datasets, in combination with other facilities in other spectral bands, have led to the construction of relatively large samples of galaxy groups which could be investigated for the first time from a multi-wavelength perspective. In this picture, the interaction between the thermal emission from the IGrM and the non-thermal emission, because of its links with AGN feedback, has especially been under the spotlight.

\citet{Giacintucci_2011} first presented the analysis of 18 X-ray selected galaxy groups studied through uGMRT. They found radio-filled X-ray cavities in half of the sample, and showed that the interaction between AGN and IGrM can be manifested also through confinement of fading radio lobes, changes in the morphology of AGN and variations in its spectral properties. The Complete Local Volume Groups Sample (CLoGS, \citealt{O'Sullivan_2017}) constitutes one of the first statistically complete samples of galaxy groups that have been studied in the context of the relation between AGN and IGrM. The radio analysis \citep{Kolokythas_2018} of this optically selected sample of 53 systems has revealed a high detection rate ($\sim92\%$) of radio sources in group dominant galaxies. Furthermore, they found that jetted AGN are more frequently detected in X-ray bright groups, and that radio non-detections are usually coincident with X-ray faint systems. Finally, their analysis suggests that central AGN in groups may inject energy in the surrounding environment through powerful periodical bursts, as AGN heating sometimes overcomes the radiative losses from IGrM cooling. By studying the interactions between radio-loud AGN and IGrM, \citet{Ineson_2013, Ineson_2015} reported a correlation between the X-ray emission of the group and the radio power of central BGGs, which is especially prominent for AGN in a phase of radiatively inefficient accretion (i.e. LERGs). \citet{Pasini_2020} confirmed the existence of this correlation studying a sample of $\sim$250 galaxy groups through MeerKAT and JVLA survey observations, in combination with \textit{Chandra} and XMM-Newton. 

The advent of new observations from the first data release of eROSITA significantly increased the number of known galaxy groups, especially in the Southern Sky. \citet{Pasini_2022} investigated the link between AGN and ICM/IGrM in a sample of more than 500 galaxy clusters and groups detected in the eFEDS field, combining eROSITA with dedicated LOFAR observations. They found that BGGs have an higher chance to host a radio-loud AGN if they lie close to the group centre, while radio-quiet BGGs can show relatively large offsets. They investigated the radio/X-ray correlation mentioned above at the extremely low frequency of LOFAR, and reported an overall balance between the total kinetic output of central AGN and IGrM cooling. A trend for stronger radio galaxies always being located close to the group centre was also reported in \citet{Pasini_2021} by analysing a sample of $\sim$1000 galaxies distributed in $\sim$80 galaxy groups, together with correlations between BGG radio power and thermodynamical properties of the host group such as temperature and X-ray luminosity. 

In summary, all these studies consistently point to the crucial role of radio-loud AGN in shaping the thermodynamical state of the IGrM. The observed correlations between radio and X-ray properties across diverse samples highlight that AGN feedback operates efficiently even in the relatively shallow potential wells of galaxy groups. These results have consolidated the view that groups represent key laboratories to study the interplay between non-thermal and thermal components of the baryonic cycle, providing fundamental insights into how AGN activity regulates gas cooling and galaxy evolution in low-mass systems.

\section{Recent progress on non-thermal emission in peculiar galaxy groups}

\begin{figure}[t!]
    \centering
	\includegraphics[scale=0.21]{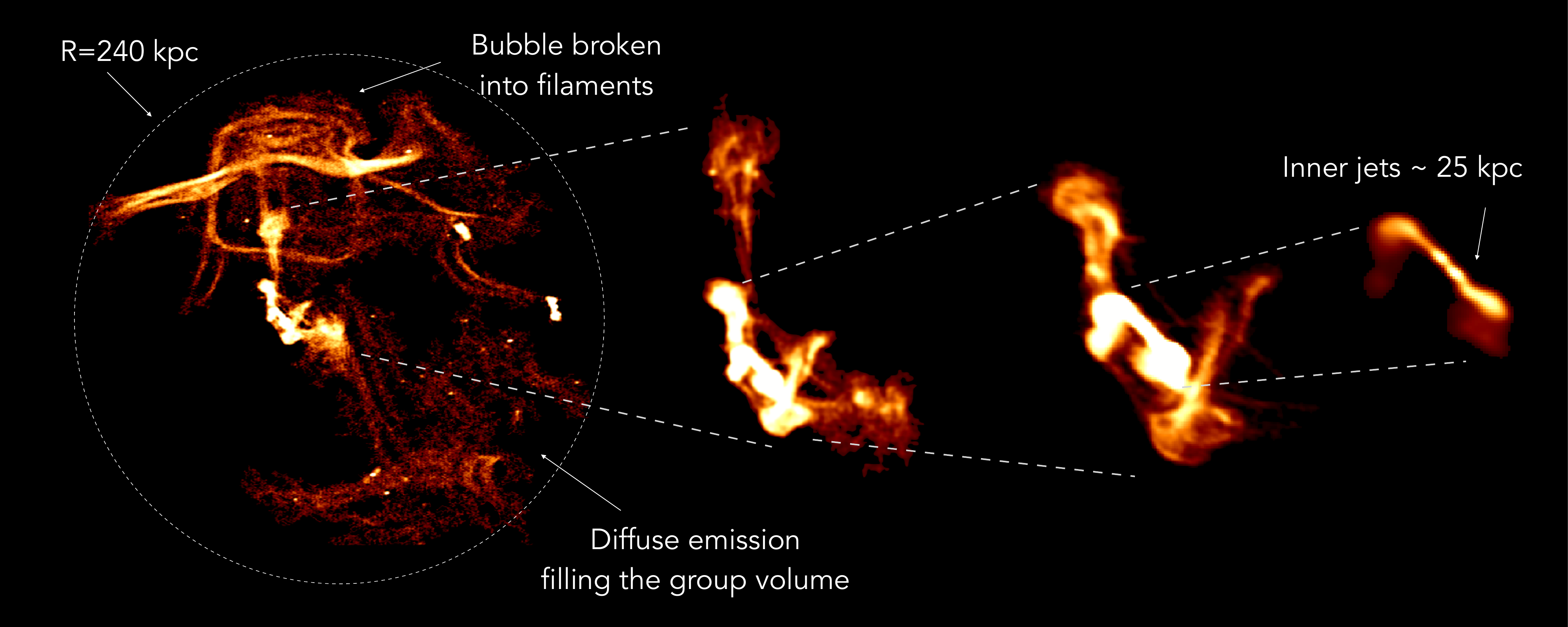}
    \caption{Multiple generations of AGN bubbles produced by the BGG of the galaxy group Nest200047 as observed by LOFAR (150 MHz) and uGMRT (400 and 700 MHz). The image is adapted from \citet{Brienza_2025}.}
    \label{fig:nest}
\end{figure}

One particularly novel opportunity offered by these new facilities, especially through low-frequency (< 1 GHz) and high sensitivity observations, is the ability to reveal previously undetected remnant plasma even within galaxy groups, i.e. the old plasma left in the environment by the jets when they switch off. 

In contrast to galaxy clusters, where the higher density of the ICM is expected to confine the old AGN bubbles, in galaxy groups these structures are instead thought to be overpressurised relative to their surroundings. This implies that they will tend to expand and fade much more rapidly than in clusters. For this reason, especially in galaxy groups, having access to low frequency data, which are sensitive to the oldest population of radio-emitting particles, is essential to track this component. It is important to stress that, even after fading below detectability, they are still expected to exert pressure on the surrounding medium, thereby providing a gentle, long-term form of feedback \citep[e.g.,][]{Bourne_2021, Stewart_2025}.

Recent observations have revealed an increasing number of detections of such aged bubbles in galaxy groups — both in well-known systems and in newly identified ones — allowing us to observe the stage when the plasma is about to mix with the surrounding medium and its energy becomes fully thermalised. In particular, observations have for the first time provided direct evidence that:

\begin{enumerate}

\item The old AGN plasma can be transported away from its original position by the motions of the intragroup medium driven by the large-scale dynamics of the system, such as sloshing motions \citep{Kolokythas_2020, Brienza_2022, Shulevski_2024};

\item The old AGN plasma can extend over large portions of a galaxy group, reaching distances up to R500 \citep[e.g.,][]{Brienza_2021, Rajpurohit_2025}. This implies that potentially the volume of galaxy groups may be filled with a non-thermal plasma component, as shown in Fig. \ref{fig:nest};

\item The old AGN bubbles can fragment into thin (a few kpc) and long (up to hundreds of kpc) filamentary structures, possibly in the presence of a turbulent environment. These filaments are likely supported by magnetic fields, which can act against instabilities and slow down the complete mixing of the non-thermal plasma with the external medium and the full thermalisation of its energy. Indeed, as the resolution and sensitivity of modern instruments improve, it is becoming clear that filamentary structures are often present in non-thermal emission, from radio galaxy lobes, to radio relics, to isolated filaments in the ICM/IGrM \citep[e.g.,][]{Owen_2014, Brienza_2021, Rudnick_2022, Churazov_2025}. This systems offer therefore key opportunities to study magnetic fields and cosmic-ray evolution, particularly in galaxy groups where they seem to be relatively common.

\end{enumerate}

The frequency of these occurrences, and how it scales with the mass and dynamical state of the system, remains, however, to be investigated. Particularly interesting are systems where multiple bubbles associated with the BGG are detected simultaneously, as this allows direct estimates of the jet duty cycle through spectral ageing modelling. To date, this type of analysis has been performed for only a handful of sources \citep[e.g.,][]{Candini_2023, Brienza_2022, Brienza_2025}, as it requires long observing campaigns with multiple instruments to cover a sufficient portion of the radio spectrum at matched sensitivity and \textit{uv}-coverage. Results from these case studies suggest jet active phases lasting a few tens of Myr, interleaved with quiescent periods ranging from a few Myr up to $\sim$100 Myr. Expanding the systematic study of jet duty cycles in galaxy groups in the SKA era will be crucial for quantifying the cumulative energy released by jets in these systems.

Finally, studying the amount and evolution of the old non-thermal plasma in galaxy groups is important for understanding how diffuse radio sources, such as halos and relics, form in merging galaxy clusters \citep[e.g.][]{vanWeeren_2019}. Observations indicate that the power of these radio sources strongly depends on the mass of the system \citep[e.g.][]{deGasperin_2014,Duchesne_2021d,Cuciti_2021}, consistent with the idea that a fraction of the total merger energy (up to $\sim10^{64}$ erg, \citealt{Ferrari_2008, Brunetti_2014}) is converted into turbulence and shocks, which then generate the observed radio emission. Most halos and relics are indeed found in clusters more massive than $5 \times 10^{14}~M_{\odot}$.

Although mergers between galaxy groups release at least an order of magnitude less energy—making turbulence and shocks less efficient—recent sensitive observations have significantly increased the number of detections in lower-mass clusters ($10^{14} \lesssim {\rm M}_{\odot} \lesssim 5\times10^{14}$, e.g. \citealt{Knowles_2016, deGasperin_2017, Kale_2017, Dwarakanath_2018, Paul_2021, Hoang_2021, Botteon_2021, Rajpurohit_2025}) opening a new observational regime that will only be fully accessible with the SKA.

Even without clear detections of halo/relic-like sources in galaxy groups, studying their AGN-driven non-thermal content remains crucial to get a grasp of the non-thermal budget of galaxy clusters \citep[e.g.,][]{Vazza_2023, Vazza_2025}. This insight is indeed key to understanding the formation of diffuse radio sources in more massive systems, which, according to the hierarchical structure formation model, are assembled from these smaller building blocks.\\

\section{Magnetic fields in galaxy groups}

Due to the hot, ionised nature of the IGrM and the ICM, it is acknowledged that magnetic fields play a crucial role. For example, they act to influence the transport of cosmic rays and conduction of heat in the medium, and thus are critical in balancing the energy budget of the IGrM/ICM \citep[e.g.][]{Rephaeli_2008,Drake_2021}. In galaxy clusters, the non-thermal pressure is measured to be of the order of $\sim 6\% - 10\%$ \citep[e.g.][]{Eckert_2019} making accurate quantification of the magnetic field properties of groups and clusters also crucial for precision cosmology considerations.

In quantifying magnetic field topography in groups we face many challenges. Precision reconstruction typically involves use of Faraday Rotation Measure (RM) synthesis in so-called `RM-grid' studies, which requires a large number of embedded or background linearly-polarised sources. While this has been achieved for several clusters \citep[e.g.][]{Bonafede_2010, Bonafede_2013, Pagliotta_2025} and on inter-cluster scales \citep[e.g.][]{Balboni_2023} the same largely cannot be said for groups, which require far higher source densities than are typically achievable with previous generation instrumentation. One notable example is the Fornax cluster, which is sufficiently nearby to allow for detailed RM grid studies \citep[e.g.][]{Anderson_2021, Loi_2025} despite being a poor cluster with a mass of $M_{\rm vir} \sim 6 \times 10^{13} ~ {\rm M_{\odot}}$ and thus straddles the boundary of groups and clusters.

For lower-mass groups with fewer member galaxies, we are often limited to individual studies of resolved polarised emission \citep[e.g.][]{NikielWroczynski_2020,Riseley_2025} which can suffer from degeneracies in fitted parameters, and inaccuracies in our modeling and/or assumptions, limiting the insights that they can provide. However, dedicated wide-deep polarisation surveys have demonstrated that RM grid studies can be powerful at probing the magnetic properties of the IGrM \citep[e.g.][]{Anderson_2024}.

\section{The role of SKA}

The SKA pathfinders and precursors discussed in the previous sections demonstrate both the scientific potential and the technical feasibility of deep, wide-area radio surveys targeting galaxy groups, but their sensitivity and resolution still constitute limiting factors for a complete census of relatively low-power AGN, remnant plasma, and diffuse synchrotron structures in the IGrM. Current radio surveys such as e.g. LoTSS \citep{Shimwell_2022} and MIGHTEE \citep{Jarvis_2016}, together with uGMRT deep fields and targeted JVLA imaging, have enabled the construction of samples including $\sim$ hundreds of galaxy groups (see Sec. \ref{sec:statistics}), but are primarily limited by either survey area, surface-brightness sensitivity and/or short-baseline coverage (depending on the instrument).

SKA represents the next step for our understanding of non-thermal emission in galaxy groups: in its AA4 configuration, SKA-Low will deploy 512 stations over baselines up to $\sim$75 km, achieving resolutions of $\sim 3''$ at 300 MHz and excellent surface-brightness sensitivity for diffuse, low-frequency emission, enabling deep low-frequency mapping of aged synchrotron plasma. SKA-Mid in AA4 configuration will instead provide $\sim 0.3 - 0.5''$ resolution at 1.4 GHz, deploying 197 antennas and achieving $\sim \mu$Jy/beam sensitivity in $\sim$1 hour. These technical specifications are directly aligned with the needs of galaxy group science. For instance, low-power sources with $L_{1.4 \rm GHz}\sim 10^{21} - 10^{22}$ W Hz$^{-1}$ (the regime expected to dominate mechanical feedback in groups) would potentially become detectable out to $z\sim1$ in $\sim$1 hour per pointing surveys; this sensitivity, together with nearly continuous frequency coverage from 50 MHz to 15 GHz (SKA-Low + SKA-Mid), enables robust spectral–ageing classification (e.g. active/remnant/fossil) and dramatically increases the completeness of diffuse, low-surface-brightness detections compared to the pathfinder surveys mentioned above. By resolving compact cores, jets, and lobes, SKA will measure the duty cycle and energetics of AGN feedback across cosmic time, constraining the recurrence and lifetime of jet outbursts and quantifying their energy coupling to the IGrM. The broad frequency coverage enables spectral index mapping throughout an unprecedented frequency range, both in-band and through the combination of SKA-Low and SKA-Mid. This is critical for spectral ageing analyses of radio lobes and remnant plasma, allowing SKA to disentangle recent AGN outbursts from remnant or re-accelerated plasma. By measuring spectral curvature and break frequencies thanks to the low-frequency coverage of SKA-Low, it will be possible to reconstruct the energy injection history and radiative ages of jets and lobes, directly linking radio properties to feedback cycles. Spatially resolved spectral-ageing maps across thousands of groups would enable observers to convert observed break frequencies into radiative ages \citep[e.g.,][]{Harwood_2013, Harwood_2015} and, in turn, constrain whether heating is dominated by frequent low-power episodes or rare powerful outbursts. Furthermore, it would be possible to perform robust measurements of duty cycles (via remnant/active fractions and spectral-curvature demographics).

The unmatched sensitivity of SKA to diffuse, low-surface-brightness emission will also be a key factor for galaxy group studies. The dense core of SKA-Low\footnote{Largest Angular Scale $>$110' at bandwidth center.} and the excellent short-baseline coverage of SKA-Mid\footnote{Largest Angular Scale $>$450' at bandwidth center.} make them uniquely capable of detecting extended ($>$100 kpc) synchrotron emission, such as radio mini-halos, buoyant lobes, and merger-induced relics in group environments. Such features, which typically have very low surface brightness, are beyond the reach of current surveys. Mapping these structures in large samples will provide direct measurements of jet-driven cavity sizes, buoyancy timescales, and energy content, enabling precise estimates of AGN mechanical power and its role in offsetting radiative cooling in group cores. Combined with complementary X-ray observations (e.g., from eROSITA), SKA data will deliver robust constraints on the thermodynamic balance of the IGrM. 

Quantitatively, by integrating the halo mass function of \citet{Tinker_2008}, we obtain an estimate of $\sim 10^6 - 10^7$ halos\footnote{This was done through {\ttfamily cluster\_toolkit (\url{https://cluster-toolkit.readthedocs.io/en/latest/source/massfunction.html})}.} in the SKA sky with masses in the range $10^{13} - 10^{14}$ M$_\odot$ out to $z \sim 1$. Given that a fraction of them would likely not host any radio-bright AGN, and assuming an active fraction of $\sim$10\% \citep[e.g.,][]{Best_2005, Best_2007}, this yields an estimate of $\sim 10^{4} - 10^{5}$ systems. Increasing to this point the number of detected group-central radio AGN systems would enable the first population-level determination of the kinetic luminosity function. The unmatched combination of sensitivity, resolution, survey speed and bandwidth will convert current evidence (largely based on single-target studies) for jet-driven heating in nearby groups into a quantitative, population-level science: measuring kinetic power distributions, duty cycles, episode durations, spectral-age demographics and magnetic-field properties across $10^{4} - 10^{5}$ groups in the $10^{13} - 10^{14}$ M$_\odot$ regime is essential to reach the statistical leverage required to robustly link AGN microphysics to galaxy evolution in the dominant cosmic environment of galaxy groups.

\begin{figure}
\centering
\includegraphics[scale=0.45]{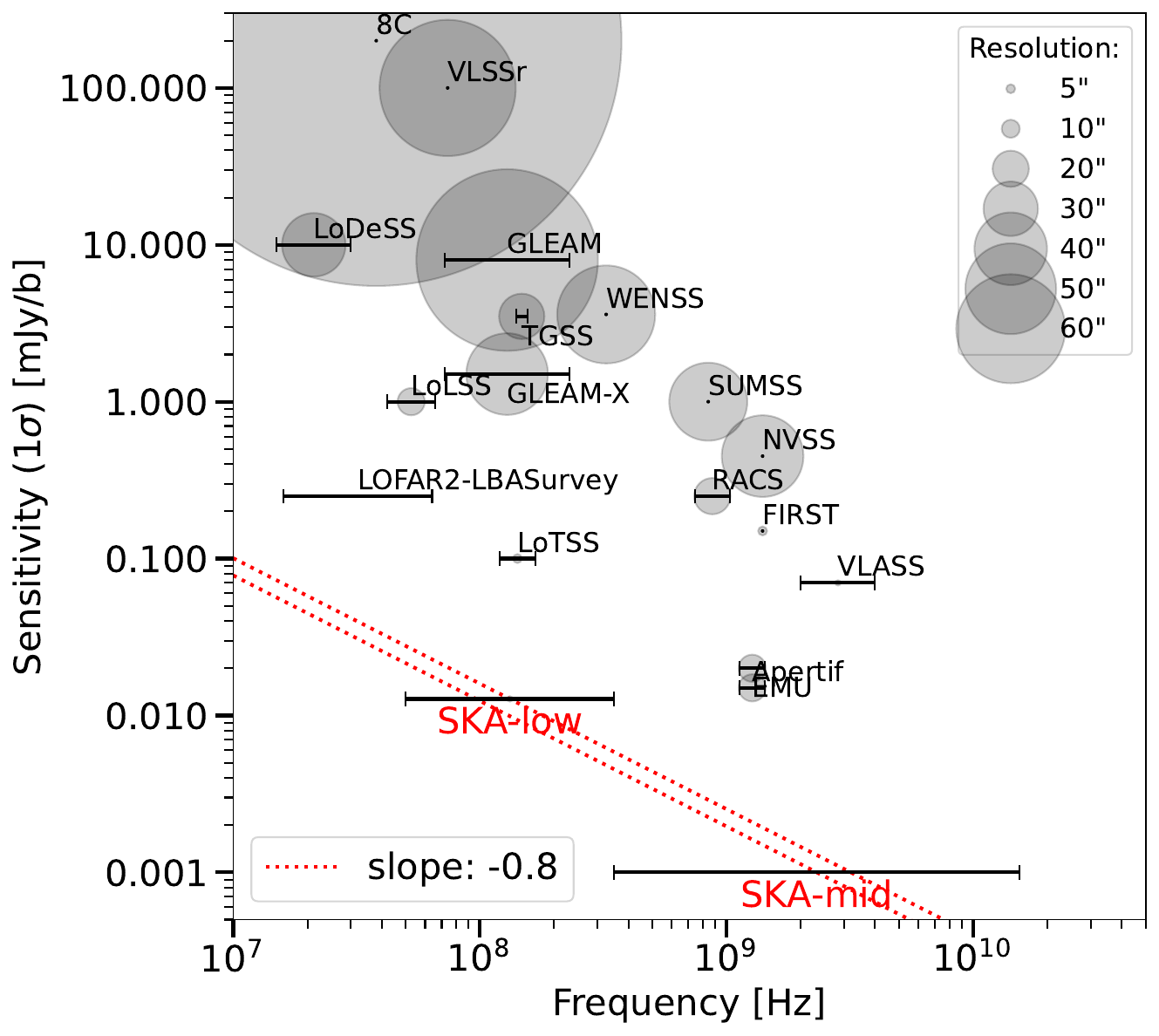}
\caption{Comparison of sensitivity, frequency coverage and resolution for the main radio continuum surveys between 10 MHz and 15 GHz. In red we show the potential sensitivity and frequency coverage of hypothetic surveys ($\sim$1 hour per pointing) carried out with SKA-Low and SKA-Mid.}
\label{fig:surveys}
\end{figure}

The extreme survey speed of SKA will also represent a key factor. In Fig. \ref{fig:surveys} we show a rough estimate of the potential of SKA-Low and SKA-Mid in AA4 configuration to carry out all-sky surveys with just $\sim$1-hour integration time per pointing (per band). SKA-Low would reach \textit{rms} noise $\sim$10 $\mu$Jy/beam with a resolution of $\sim 6''$ at 300 MHz, while SKA-Mid would open a new window on the sub-$\mu$Jy/beam regime with a resolution of $\sim 0.5''$. If surveys like these were to be carried out, the improvement in terms of both sensitivity and resolution, compared to current facilities, would be of more than an order of magnitude in both frequency bands, as clearly shown in Fig. \ref{fig:surveys}. 

In more detail, the extremely good low-frequency sensitivity of a survey carried out with SKA-Low would enable to probe the properties of hundreds of thousands of galaxy groups, together with their central radio AGN, especially when in combination with other multi-wavelength instruments such as e.g. eROSITA, XMM-Newton and Euclid. On the other hand, the multi-frequency coverage and high-resolution granted by the synergy with SKA-Mid would enable to investigate the physics of peculiar group-scale non-thermal emission with a level of detail impossible to reach with current instruments. Furthermore, adding deep-field observations of relatively small sky areas would provide detailed case studies of feedback and plasma physics in individual systems, as well as probe for the first time the ultra-faint diffuse regime. We acknowledge, however, that the smaller field of view of SKA-Mid would require a significantly larger number of pointings, and therefore longer total observing times, compared to SKA-Low, to cover similar sky areas.

We can forecast SKA's impact on galaxy group RM grids by extrapolating from early science with the Pathfinders and Precursors. ASKAP's POlarisation Sky Survey of the Universe's Magnetism \citep[POSSUM;][]{Gaensler_2025} predicts RM densities of around 30 to 50 RMs per square degree at $20''$ resolution with an \textit{rms} noise around $18 ~ \mu$Jy/beam sensitivity from $\sim 10$~hour tracks. In comparison, SKA will achieve a factor $\sim 7 - 40$ improvement in resolution and $\sim 20$ in sensitivity (depending on frequency) in $\sim1$ hour on-source. As such, extrapolating polarised source counts \citep[e.g.][]{Stil_2014,Hales_2014} to SKA's $\sim \mu$Jy sensitivities we would tentatively expect RM grid densities of up to $\sim 310$ sources per square degree. Following the size expectation of the group halos, it follows that in the local Universe ($z \sim 0.02$) we might expect up to a few hundred RMs along the line of sight through low-mass halos ($10^{13} ~ M_{\odot}$); out to redshift $z \sim 0.5$ however we might only expect a few RMs per group halo. These estimates heavily depend on the poorly constrained evolution of polarised source counts in different source populations, another open question to be examined with SKA. Additionally, SKA's nearly-continuous frequency coverage from low to high frequencies will enable not just precision studies of the magnetised IGrM through dense RM grids, but also detailed studies of the magnetoionic medium through QU-fitting and the wavelength-dependent depolarisation of sources viewed through the IGrM \citep{skaloi}, extending techniques from the cluster regime \citep[e.g.][]{Osinga_2022,Rajpurohit_2022,Pasetto_2026}.

In conclusion, a combined strategy exploiting a wide, short survey and deep smaller fields would maximize the scientific return on duty-cycle, energetics and feedback-efficiency measurements, as recently taught by LOFAR surveys \citep[e.g.,][]{Shimwell_2022, deGasperin_2023} and deep fields \citep[e.g.,][]{Williams_2021, Sabater_2021, Best_2023}. Together, these datasets will enable the first statistical characterization of AGN feedback in galaxy groups across an unprecedentedly large range of mass and redshift, linking the microphysics of jets and lobes to the macroscopic evolution of group baryon cycles. We stress that surveys carried out through pathfinder facilities such as LOFAR, MeerKAT, and ASKAP, while providing critical preparatory datasets, are limited to $\sim$5–10 times (depending on the instrument) higher noise levels and worse resolutions at similar frequencies, as shown in Fig. \ref{fig:surveys}. SKA will thus bridge the gap between the detailed studies achievable in the very nearest groups and the need for a comprehensive, cosmologically significant census of AGN-driven feedback in the most common environment for galaxies in the Universe. 
On a final note, we emphasise that, in addition to the BGG, the sensitivity and angular resolution of SKA will make it possible to extend feedback studies to the full population of group member galaxies. In particular, SKA will enable the detection and characterisation of faint jets, lobes and relic plasma associated with AGN hosted by satellite galaxies, allowing their cumulative and individual contributions to the non-thermal energy budget of the IGrM to be quantified. By combining this information with the diffuse, low-surface-brightness mapping of aged plasma on group scales, it will be possible to investigate whether feedback in groups is dominated by a single central engine, or instead arises from repeated, distributed episodes of AGN activity across multiple members. This approach will allow a more complete, population-level assessment of how AGN feedback operates and affect the group environment and its evolution.

\section{Conclusions}

In this chapter, we have reviewed recent advances in our understanding of AGN feedback in galaxy groups, and identified the outstanding questions that future observations with the SKA will be able to address. The impact of feedback in groups is enhanced by their shallow potentials and by their key role in structure formation. Therefore, understanding feedback in groups is crucial to understand in turn galaxy evolution, for explaining observed deviations from self-similar scaling relations, and for modeling the baryon content and star formation in these systems. Non-thermal radio emission plays a key role in this sense, as it traces the outburst history of AGN, provides a direct probe of the energy released by SMBHs, and is essential to understand how this energy couples to the surrounding IGrM. Moreover, spectral and morphological analyses of non-thermal emission enable to estimate the radiative ages of relativistic electrons, constrain magnetic field and particle acceleration mechanisms, and thus reconstruct the duty cycle, efficiency, and time evolution of AGN feedback within the group environment.

Historically, the study of feedback in these environments has been hampered by their low surface brightness (e.g. compared to more massive galaxy clusters), especially in the X-ray band, which makes it harder to investigate the link between the non-thermal emission and the group environment. Furthermore, current radio instruments are not always able to provide the combination of resolution and sensitivity necessary to study statistically large samples. This chapter has discussed how the advent of SKA promises to revolutionize this topic. In particular, the main scientific questions that we have highlighted, and for which SKA will represent a crucial step forward, include:

\begin{itemize}
\item What are the duty cycle and kinetic energy output of AGN in group-central galaxies, and how do they regulate the cooling and star formation in the IGrM?
\item How common are aged or remnant radio lobes and other diffuse/extended non-thermal structures in galaxy groups, and what role do they play in heating and stirring the IGrM?
\item What is the kinetic luminosity function of radio AGN in galaxy groups, and how does their power distribution evolve over cosmic time?
\item Are galaxy group cores maintained by frequent low-power AGN outbursts or by rare, powerful episodes of feedback?
\end{itemize}

Thanks to its unparalleled combination of sensitivity, resolution, and frequency coverage, SKA will make it possible to tackle these questions directly. In particular, SKA-Low and SKA-Mid will be able to detect very faint, steep-spectrum emission and resolve compact jet and lobe structures across a wide frequency range. Potentially, all-sky surveys carried out with SKA would be able to expand the sample of known group-central radio AGN from the hundreds to $\sim10^4$–$10^5$ systems. This will enable the first population-level measurement of the AGN kinetic luminosity function and the distribution of feedback energetics, as well as robust spectral-ageing analysis of radio lobes over thousands of groups. In summary, SKA promises to tackle many of the key questions related to AGN feedback in galaxy groups identified above. Such studies will yield a comprehensive, quantitative picture of how radio-mode feedback operates in the most common environment for galaxies, and will clarify the essential role that AGN feedback plays in galaxy evolution.

\bibliographystyle{abbrvnat-maxbibnames4.bst}
\bibliography{chapter} 

\end{document}

%% file: journal-names.tex
\newcommand{\actaa}{Acta Astron.} 
\newcommand{\araa}{Annu. Rev. Astron. Astrophys.} 
\newcommand{\aar}{Astron. Astrophys. Rev.} 
\newcommand{\ab}{Astrobiol.} 
\newcommand{\aj}{Astron. J.} 
\newcommand{\apj}{Astrophys. J.} 
\newcommand{\apjl}{Astrophys. J. Lett.} 
\newcommand{\apjs}{Astrophys. J. Suppl. Ser.} 
\newcommand{\ao}{Appl. Opt.} 
\newcommand{\apss}{Astrophys. Space Sci.} 
\newcommand{\aap}{Astron. Astrophys.} 
\newcommand{\aapr}{Astron. Astrophys. Rev.} 
\newcommand{\aaps}{Astron. Astrophys. Suppl.} 
\newcommand{\baas}{Bull. Am. Astron. Soc.} 
\newcommand{\caa}{Chinese Astron. Astrophys.} 
\newcommand{\cjaa}{Chinese J. Astron. Astrophys.} 
\newcommand{\cqg}{Class. Quantum Gravity} 
\newcommand{\gal}{Galaxies} 
\newcommand{\gca}{Geochim. Cosmochim. Acta} 
\newcommand{\icarus}{Icarus} 
\newcommand{\jcap}{J. Cosmol. Astropart. Phys.} 
\newcommand{\jgr}{J. Geophys. Res.} 
\newcommand{\jgrp}{J. Geophys. Res.: Planets} 
\newcommand{\jqsrt}{J. Quant. Spectrosc. Radiat. Transf.} 
\newcommand{\memsai}{Mem. Soc. Astron. Italiana} 
\newcommand{\mnras}{Mon. Not. R. Astron. Soc.} 
\newcommand{\nat}{Nature} 
\newcommand{\nastro}{Nat. Astron.} 
\newcommand{\ncomms}{Nat. Commun.} 
\newcommand{\nphys}{Nat. Phys.} 
\newcommand{\na}{New Astron.} 
\newcommand{\nar}{New Astron. Rev.} 
\newcommand{\physrep}{Phys. Rep.} 
\newcommand{\pra}{Phys. Rev. A} 
\newcommand{\prb}{Phys. Rev. B} 
\newcommand{\prc}{Phys. Rev. C} 
\newcommand{\prd}{Phys. Rev. D} 
\newcommand{\pre}{Phys. Rev. E} 
\newcommand{\prl}{Phys. Rev. Lett.} 
\newcommand{\psj}{Planet. Sci. J.} 
\newcommand{\planss}{Planet. Space Sci.} 
\newcommand{\pnas}{Proc. Natl Acad. Sci. USA} 
\newcommand{\procspie}{Proc. SPIE} 
\newcommand{\pasa}{Publ. Astron. Soc. Aust.} 
\newcommand{\pasj}{Publ. Astron. Soc. Jpn} 
\newcommand{\pasp}{Publ. Astron. Soc. Pac.} 
\newcommand{\rmxaa}{Rev. Mexicana Astron. Astrofis.} 
\newcommand{\sci}{Science} 
\newcommand{\sciadv}{Sci. Adv.} 
\newcommand{\solphys}{Sol. Phys.} 
\newcommand{\sovast}{Soviet Ast.} 
\newcommand{\ssr}{Space Sci. Rev.} 
\newcommand{\uni}{Universe} 